\documentclass[preprint,showpacs,showkeys,aps,prb]{revtex4}
\usepackage[english]{babel}

\usepackage[dvips]{graphicx}
\usepackage{amssymb}

\begin{document}


\title{Tensor Temperature and Shockwave Stability \\
in a Strong Two-Dimensional Shockwave. }

\author{Wm. G. Hoover and Carol G. Hoover \\
Ruby Valley Research Institute \\ Highway Contract 60,
Box 598, Ruby Valley 89833, NV USA}

\date{\today}




\vskip 0.5cm

\begin{abstract}

The anisotropy of temperature is studied here in a strong two-dimensional
shockwave, simulated with conventional molecular dynamics.  Several forms
of the kinetic temperature are considered, corresponding to different
choices for the local instantaneous stream velocity.  A local
particle-based definition omitting any ``self'' contribution to the
stream velocity gives the best results.  The configurational
temperature is not useful for this shockwave problem. Configurational
temperature is subject to a shear instability and can give local negative
temperatures in the vicinity of the shock front.  The decay of 
sinusoidal shockfront perturbations shows that strong two-dimensional
planar shockwaves are stable to such perturbations.

\end{abstract}

\maketitle

\section{Introduction}

Shockwaves are useful tools for the understanding of material behavior far
from equilibrium\cite{b1,b2,b3,b4}.  The high-pressure physicist Percy Bridgman
played a key role in the adaptation of experimental shockwave physics to
the thermodynamic characterization of materials at high pressure\cite{b4}.
Because shockwaves join two purely equilibrium
states, shown to the left and right of the central shockwave
in Figure 1, the experimental and computational difficulties
associated with imposing  nonequilibrium boundary conditions are absent.

Begin by assuming that the flow is both stationary and one-dimensional.
Such a flow gives
conservation of the mass, momentum, and energy fluxes throughout the
shockwave.  Thus the fluxes of mass, momentum, and energy,
$$
\rho u, P_{xx} + \rho u^2, \rho u[e + (P_{xx}/\rho ) + (u^2/2)] + Q_x \ ,
$$
are constant throughout the system, even in the far-from-equilibrium
states within the shockwave.  Here $u$ is the flow velocity, the
velocity in the $x$ direction, the direction of propagation.
We use conventional notation here, $\rho$ for density, $P$ for
the pressure tensor, $e$ for the internal energy per unit mass, and
$Q_x$ for the component of heat flux in the shock direction, $x$.  It
is important to recognize that both the pressure tensor and the heat
flux vector are measured in a local coordinate frame moving with the
local fluid velocity $u(x)$.

In a different ``comoving frame'', this time moving with the shock
velocity and centered on the shockwave, cold material enters from
the left, with speed $u_s$ (the ``shock speed'') and hot material exits
at the right, with speed $u_s - u_p$, where $u_p$ is the ``piston
speed''.  Computer simulations of shockwaves, using molecular dynamics, have
a history of more than 50 years, dating back to the development of fast
computers\cite{b5,b6,b7,b8,b9,b10,b11}.  Increasingly sophisticated high-pressure
shockwave experiments have been carried out since the Second World
War\cite{b12}.

\begin{figure}
\vspace{1 cm}
\includegraphics[height=10cm,width=2cm,angle=-90]{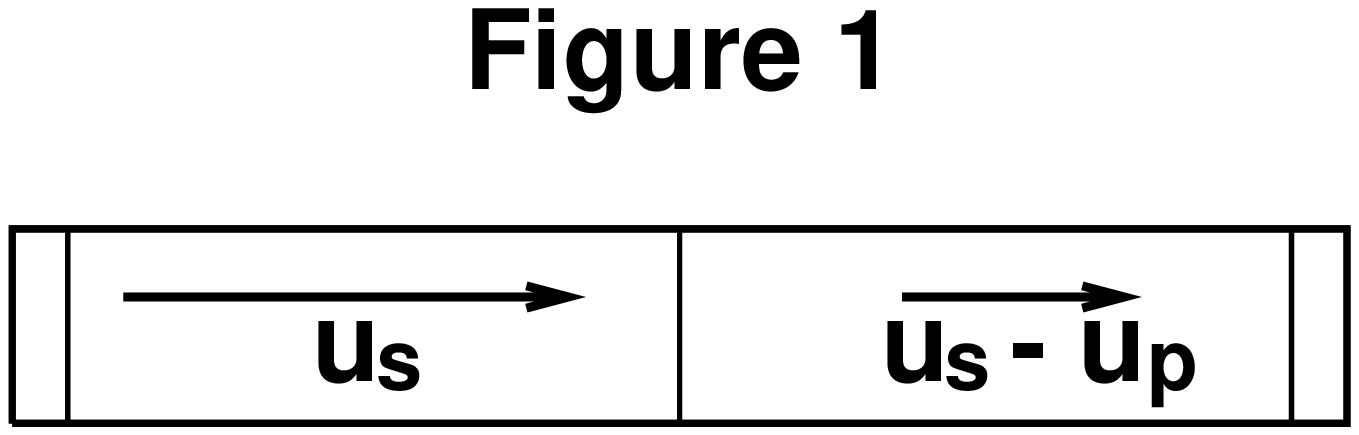}
\caption{
A stationary one-dimensional shockwave.  Cold material enters at the left
with a speed $u_s$, passes through the shockfront which separates the cold
material from the hot, and exits at the right with speed $u_s - u_p$.  The
cold-to-hot conversion process is irreversible and corresponds to an
overall entropy increase.
}
\end{figure}

In laboratory experiments it is convenient to measure the two speeds, $u_s$
and $u_p$.  These two values, together with the initial ``cold'' values of
the density, pressure, and energy,  make it possible to solve the three
conservation equations for the ``hot values'' of $\rho$, $P_{xx}$, and $e$.
A linear relation between $P_{xx}(x)$ and $V(x)=1/\rho $, results when the
mass flux $M \equiv \rho u$ is substituted into the equation for momentum
conservation:
$$
P_{xx}(x) + M^2/\rho (x) =
P_{\rm hot} + (M^2/\rho_{\rm hot}) = P_{\rm cold} + (M^2/\rho_{\rm cold}) \ .
$$
This nonequilibrium pressure-tensor relation is called the
``Rayleigh Line''.  The energy conservation relation,
$$
\Delta e = -(1/2)[P_{\rm hot} +  P_{\rm cold}]\Delta V \ ,
$$
based on equating the work of compression to the gain in internal energy,
is called the ``Shock Hugoniot Relation''.  See Figure 2 for both.  A
recent comprehensive review of shockwave physics can be found in Ref. 3.

Laboratory experiments based on this approach have detailed the equations
of state for many materials. Pressures in excess of 6TPa (sixty megabars)
have been characterized\cite{b12}.  If a constitutive relation is assumed
for the nonequilibrium anisotropic parts of the pressure tensor and heat flux,
the conservation relations give ordinary differential equations for the
shockwave profiles. With Newtonian viscosity and Fourier heat conduction
the resulting ``Navier-Stokes'' profiles have shockwidths on the order of
the mean free path\cite{b7,b8}.

\begin{figure}
\includegraphics[height=10cm,width=8cm,angle=-90]{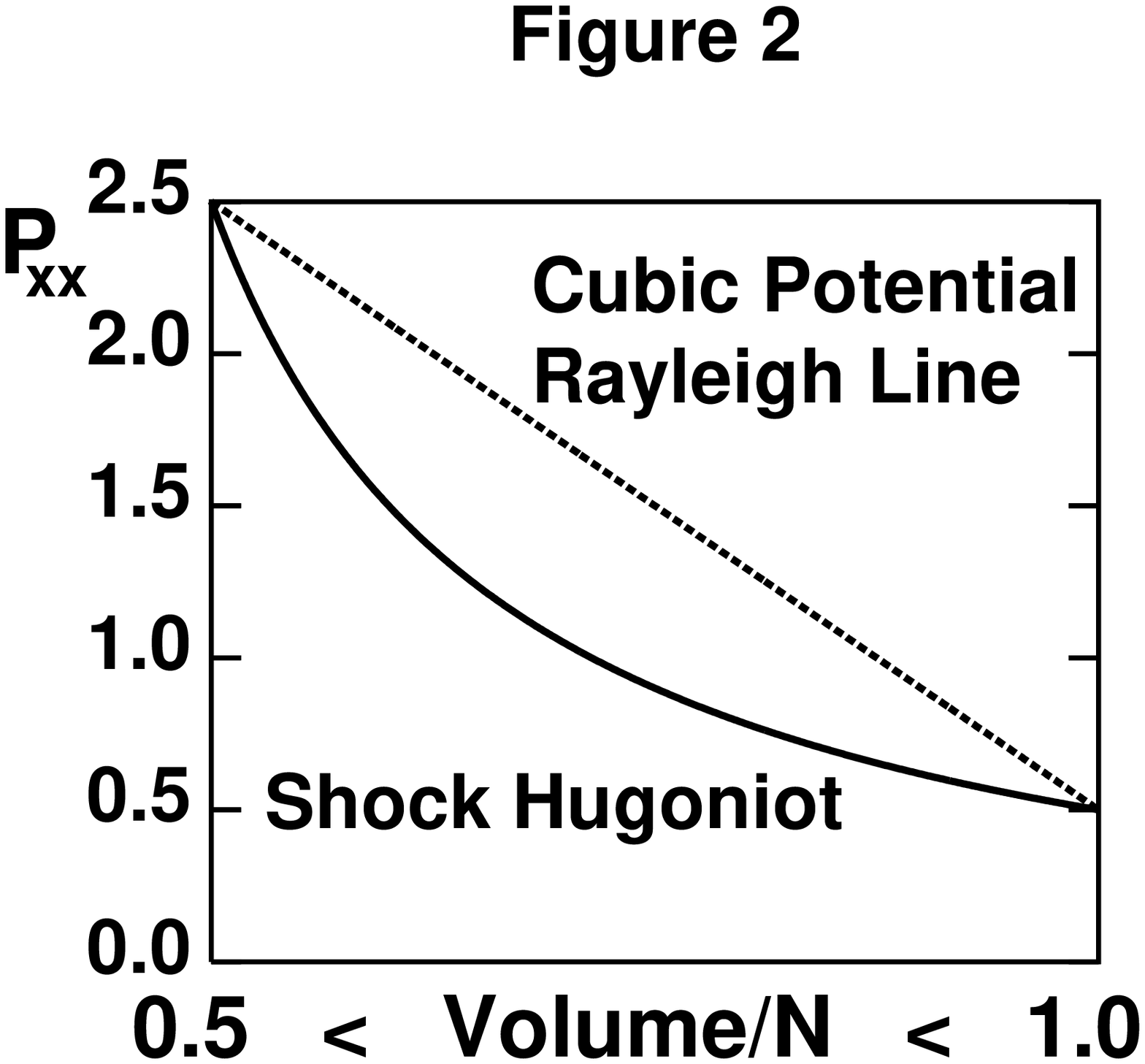}
\caption{
Calculated Rayleigh Line and Shock Hugoniot relation for the simple repulsive van
der Waals' equation described in the text.  The Rayleigh line includes
nonequilibrium states within the shock while the Shock Hugoniot line is the
locus of all equilibrium states accessible by shocking the initial state.
}
\end{figure}

Early theoretical analyses of shockwaves emphasized solutions of the
Boltzmann equation.  Mott-Smith's approximate solution of that
equation\cite{b13}, based on the weighted average of two equilibrium
Maxwellian distributions, one hot and one cold, revealed a temperature
maximum (and a corresponding entropy maximum) at the shock center, for
shocks with a Mach number exceeding two.  The Mach number is the ratio of the
shock speed to the sound speed.  Twenty years later, molecular dynamics
simulations showed shockwidths of just a few atomic
diameters.\cite{b5,b6,b7,b8}  These narrow
shockwaves agreed nicely with the predictions of
the Navier-Stokes equations for relatively weak shocks.  At higher
pressures there is a tendency for the Navier-Stokes profiles to
underestimate the shockwidth.  Some of the computer simulations have
shown the temperature maximum at the shock front predicted by
Mott-Smith\cite{b8,b10}.  These temperature maxima are constitutive
embarrassments --- they imply a negative heat conductivity for a part
of the shockwave profile.

Our interest here is twofold.  We want to check on the stability of
planar shockwaves (the foregoing analysis assumes this stability) and
we also want to characterize the anisotropy of temperature in the shock.
This latter topic is particularly interesting now in view of the several
definitions of temperature applied to molecular dynamics
simulations\cite{b8,b9,b14,b15,b16,b17,b18}.  A configurational-temperature
definition as well as several kinetic-temperature definitions can all be
applied to the shockwave problem.

The plan of the present work is as follows.  Section II describes the material
model and computational setup of the simulations.  Section III describes
the computation of the shock profiles and the analysis confirming
their stability.  Section IV compares the various kinetic and
configurational temperature definitions for a strong stable
shockwave.  Section V details our conclusions.

\section{Shockwave Simulation for a Simple Model System}

We simulate the geometry of Figure 1 by introducing equally-spaced
columns of cold particles from a square lattice.  The initial lattice
moves to the right at speed $u_s$. The interior of the system is purely
Newtonian, without any boundary, constraint, or driving forces.  Those
particles coming within the range of the forces $(\sigma = 1)$ of the
righthand boundary have their velocities set equal to $u_s - u_p$ (the
mean exit velocity) and are discarded once they reach the boundary.  This
boundary condition, because it corresponds to a diffusive heat sink at the
exit, only affects the flow in the vicinity of that boundary.
We initially chose the speeds $u_s = 2$ and $u_p = 1$ to correspond to twofold
compression.  The approximate thermal and mechanical equations of
state,
$$
e = (\rho /2) + T \ ; \ P = \rho e \ .
$$
together with the initial values,
$$
\rho _{\rm cold} = 1 \  ; \ T_{\rm cold} = 0 \ ; \ 
P_{\rm cold} = (1/2) \ ; \ e_{\rm cold} = (1/2) \ ,
$$
give corresponding ``hot'' values.  These two sets of thermodynamic data
mutually satisfy the Rayleigh Line and Shock Hugoniot Relation for
twofold compression from the cold state:
$$
\rho _{\rm hot} = 2 \  ; \ T_{\rm hot} = (1/4) \ ; \ 
P_{\rm hot} = (5/2) \ ; \ e_{\rm hot} = (5/4) \ .
$$
The mass, momentum, and energy fluxes are 2, (9/2), and 6,
respectively.  For reasons explained below it was necessary to modify
these conditions slightly, using instead $u_s = 2u_p = 1.75$.

We considered a wide variety of system lengths and widths
and found no significant difference in the nature of the results.  For
convenience we show here results for a system of length $L_x = 200$
and width $L_y = 40$.  Because the density increases by a factor of two
in the center of the system the number of particles used in this case
is about 12000.  The number varies during the simulation as new particles
enter and old ones are discarded.  The total length of the run is one
shock traversal time, $200/u_s$, though the shock is itself localized
near the center of the system and in fact moves very little in our
chosen coordinate frame.  In retrospect, the simulations could just as
well have been carried out with a much smaller $L_x$.  Because we wished
to study stability we felt it necessary to use a relatively wide system.

For a two-dimensional classical  model with a weak van der Waals'
repulsion the equation of state described above follows from the simple
canonical partition function:
$$
Z^{1/N} \propto VTe^{-\rho/2T} \ ; \
PV/NkT = [\partial \ln Z/\partial \ln V]_T \ ; \
E/NkT = [\partial \ln Z/\partial \ln T]_V \ .
$$

Our original intent was to use the smooth repulsive pair potential\cite{b19}
$$
\phi = (10/\pi \sigma^2)[1 - (r/\sigma)]^3 \ {\rm for} \ r<\sigma \
\longrightarrow
$$
$$
\langle \Phi \rangle \simeq (N\rho/2)\int_V\phi (r)2\pi rdr = N\rho /2 \ .
$$
for a sufficiently large $\sigma$ (3 or so) that the simple equation of state
was accurate.  But for large $\sigma$ the shockwidth is also so large that
detailed studies are impractical.  In the end we chose to set the range of
the forces equal to unity, so that the initial pressure is actually zero
rather than 1/2.  Nevertheless, the choice of $u_s$ = 2 is still roughly
compatible with twofold compression. Equilibrium molecular dynamics
simulations for $\sigma = 1$ give the following solution to the
conservation relations for twofold compression, from
$\rho _0 = 1$ to $\rho = 2$ with $u_s = 2u_p = 1.75$:
$$
\rho u: 1.0 \times 1.75 = 2.0 \times 0.875 = 1.75 \ ;
$$
$$
P + \rho u^2: 0.0 + 1.0 \times 1.75^2 = 1.531 + 2.0\times 0.875^2 = 3.062 \ ;
$$
$$
\rho u[e +(P/\rho) + (u^2/2)] = 1.75[0.0 + 0.0 + 1.531] 
= 1.75[0.383 + 0.766 + 0.383] = 1.75 \times 1.531 \ .
$$

We introduced a sinusoidal perturbation in the initial conditions by using
twice as many columns of particles (per unit length) to the right of a
line near the center of the system:
$$
x_{\rm shock} = 6\sin (2\pi y/L_y) \ .
$$
The time development of a system starting with this sinewave displacement
perturbation is shown in Figure 3.  The panel corresponding to the time
$t = Dt$ shows that the decay is underdamped, while the following panels
show that the planar shockwave is stable.

Figure 4 illustrates the effect of
the gradual underdamped flattening of the sinewave perturbation on the density
profile.  The propagation of the wave is followed through
five shock traversal times of the observation window used in Figure 3.

\begin{figure}
\includegraphics[height=10cm,width=8cm,angle=-90]{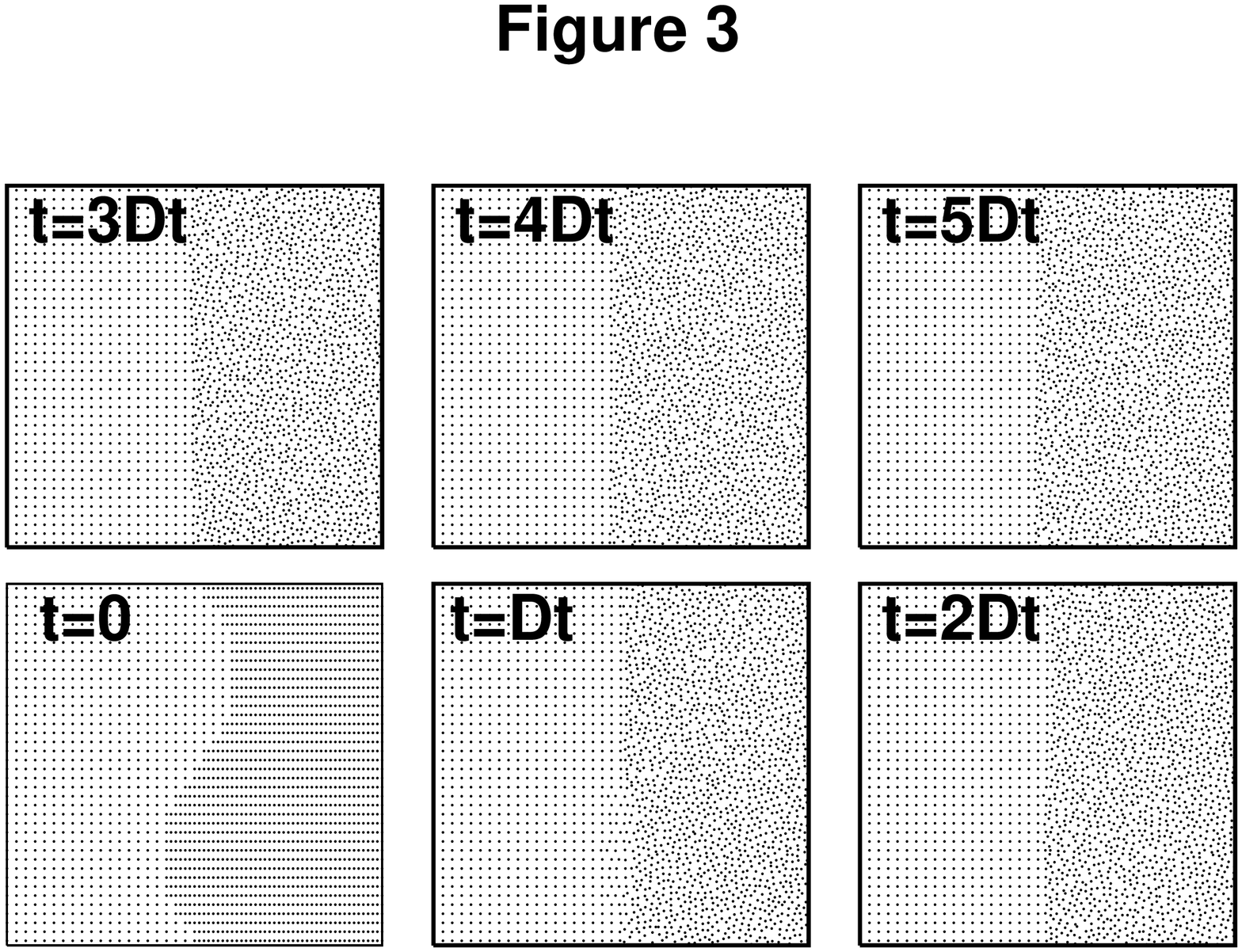}
\caption{
Particle positions in the initial condition correspond to $t=0$.  Those
to the left of the sinewave boundary move to the right at speed $u_s = 1.75$
while those to the right travel at speed $u_s - u_p = u_s/2$.  Particle
positions at five later equally-spaced times are shown too.  The time
interval $Dt = 40/u_s$ is 2000 timesteps.  The $40 \times 40$ windows
shown here would contain 1600 cold particles or 3200 hot particles at the
cold and hot densities of 1.0 and 2.0.
}

\end{figure}
\begin{figure}
\includegraphics[height=10cm,width=8cm,angle=-90]{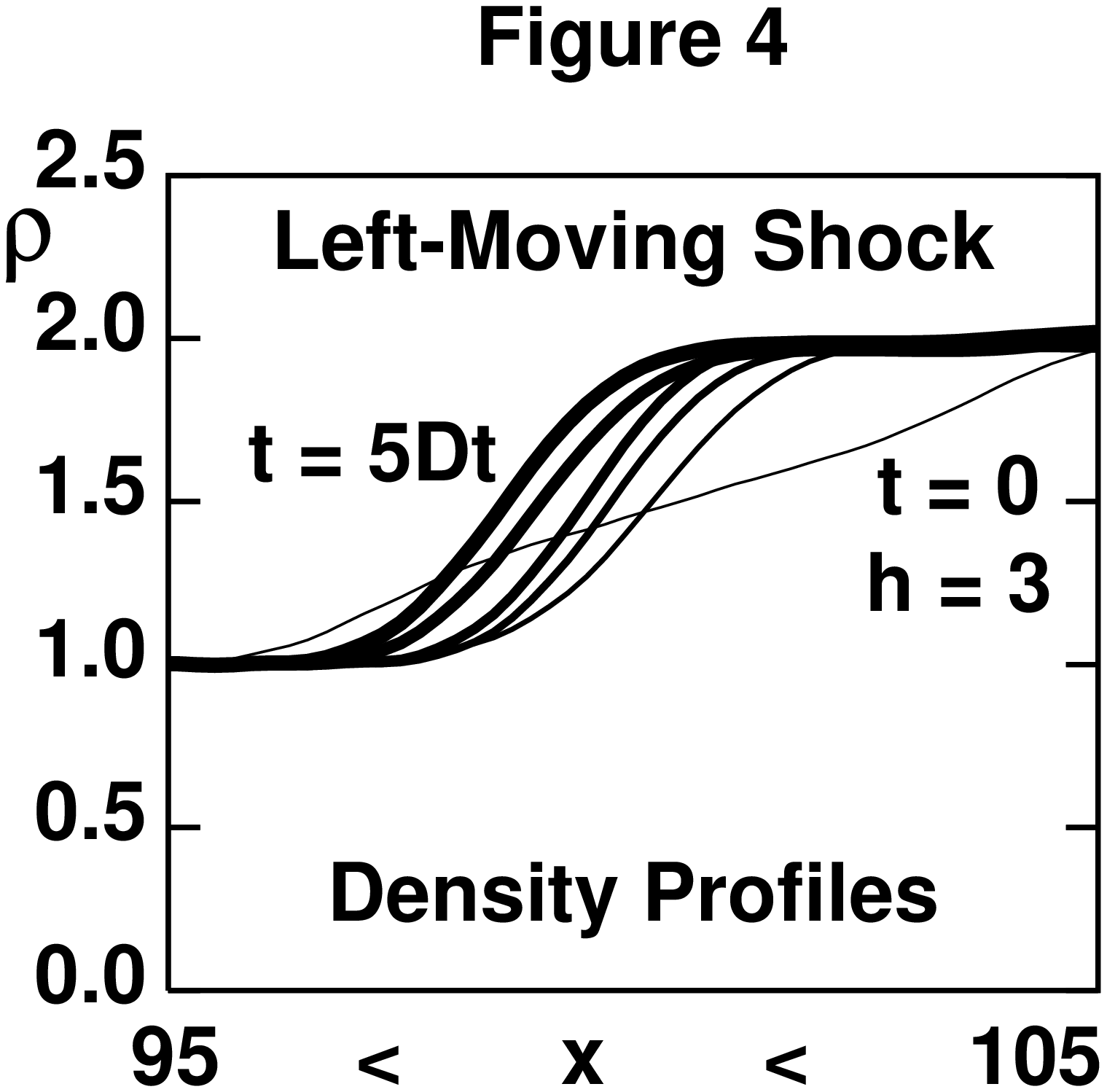}
\caption{
Density profiles at the same times as those illustrated in Figure 3.
These one-dimensional density profiles were computed with
Lucy's one-dimensional smooth-particle weight function using $h = 3$
and the particle coordinates shown in Figure 3.  Increasing time
corresponds to increasing line thickness as the shockwave moves slowly
to the left.
}
\end{figure}

The density profiles are computed with Lucy's one-dimensional weighting
function\cite{b20,b21}
$$
w^{1D}(r<h) = (5/4h)[1 - 6(r/h)^2 + 8(r/h)^3 - 3 (r/h)^4] \ ; \ r < h=3 \ 
$$
$$
\longrightarrow \int_{-h}^{+h} w(r)dr \equiv 1 \ .
$$
The density is computed by evaluating the expression
$$
\rho (x_k) = \sum w_{ik}^{1D}/L_y \ ; \ w_{ik}^{1D} \equiv
w^{1D}(|x_i - x_k|) \ ,
$$
for all combinations of Particle $i$ and gridpoint $k$ separated by
no more than $h=3$.  For comparison density profiles using a shorter
range, $h=2$, are shown in Figure 5.  These latter profiles exhibit
a wiggly structure indicating a deterioration of the averaging process.

\begin{figure}
\includegraphics[height=10cm,width=8cm,angle=-90]{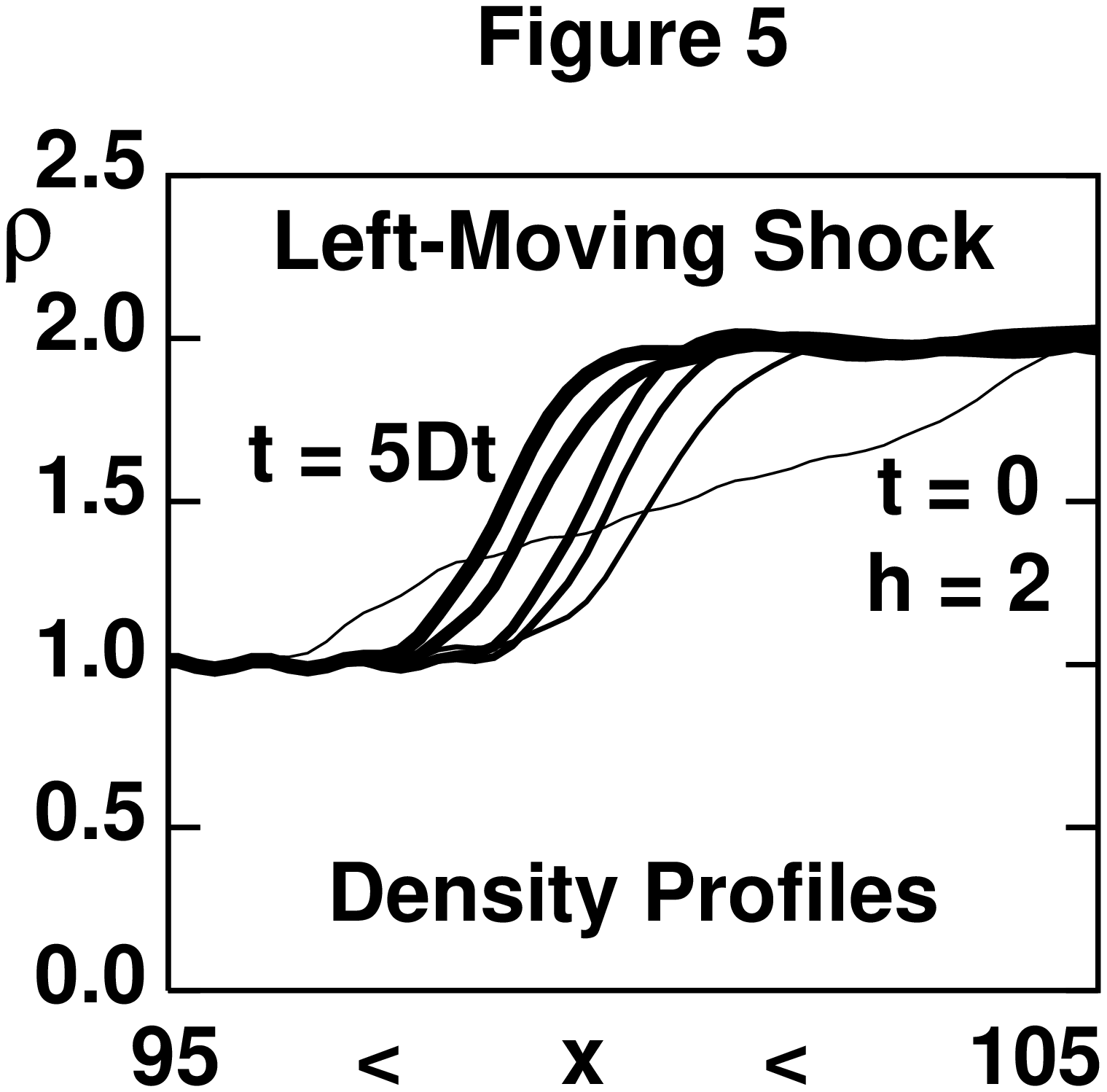}
\caption{
Density profiles exactly as in Figure 4, but with a reduced range,
$h=2$.  The shockwave moves slowly to the {\em left}.  Note the wiggly
structure, a consequence of choosing $h$ too small..
}
\end{figure}

Similar conclusions, using similar techniques, have been drawn by Hardy
and his coworkers\cite{b9,b11}, who were evidently unaware of Lucy and
Monaghan's work.  In their interesting analysis of a two-dimensional
shockwave, Root, Hardy, and Swanson use a weight function like Lucy's,
but with square
(rather than circular) symmetry and with only a single vanishing
derivative at its maximum range\cite{b11}.  They discuss the ambiguities of
determining temperature away from equilibrium and rightly conclude that
the range of the spatial weight function needs careful consideration.
Hardy\cite{b9} recently told us that his use of a spatial weighting
function was motivated by a conversation with Philippe Choquard.

The profiles we show in Figure 4 are typical, and show that the shockwidth
rapidly attains a value of about 3 particle diameters, and has no further
tendency
to change as time goes on.  A detailed study shows that the sinewave
amplitude exhibits underdamped oscillations on its way to planarity.
Evidently, for this two-dimensional problem,
the one-dimensional shockwave structure is stable.  In the next Section
we consider the temperature profiles for the stationary shockwave.

\section{Configurational and Kinetic Temperatures in the Shockwave}

The kinetic and configurational contributions to temperature and pressure
have been discussed and explored in a variety of nonequilibrium contexts.
Shockwaves, with stationary boundary conditions far from the shockfront,
allow the anisotropy of the kinetic temperature to be explored,
analyzed, and characterized with purely Newtonian molecular dynamics.
Kinetic-theory temperature is based on the notion of an equilibrium
ideal gas thermometer\cite{b22,b23,b24}. The temperature(s) measured by such a
thermometer
are given by the second moments of the velocity distribution:
$$
\{ kT_{xx}^K, kT_{yy}^K \} =
m\{\langle v_x^2 \rangle ,\langle v_y^2 \rangle \} 
$$
where the velocities are measured in the comoving frame, the frame moving
at the mean velocity of the fluid.

A dilute Maxwell-Boltzmann gas of small hard parallel cubes undergoing
impulsive collisions with a large test particle provides an explicit
model tensor thermometer for the test-particle kinetic temperature\cite{b24}.
The main difficulty
associated with kinetic temperature lies in estimating the mean
stream velocity, with respect to which thermal fluctuations define
kinetic temperature.  In what follows we compare several such
definitions, in an effort to identify the best approach.

Configurational temperature is more complicated and lacks a definite
microscopic mechanical model of a thermometer able to measure it.
Configurational temperature is based on linking two canonical-ensemble
equilibrium averages, as was written down by Landau and Lifshitz more than
50 years ago\cite{b25}:
$$
\{ kT^\Phi_{xx},kT^\Phi_{yy} \} = \{
 \langle F_x^2 \rangle / \langle \nabla ^2_x{\cal H} \rangle ,
 \langle F_y^2 \rangle / \langle \nabla ^2_y{\cal H} \rangle\} \ .
$$
${\cal H}$ is the Hamiltonian governing particle motion.  Unlike kinetic
temperature this configurational definition is independent of stream
velocity.  Its apparent dependence on rotation\cite{b19} is negligibly small
for the (irrotational) shockwave problems considered here.
We apply both the kinetic and the configurational approaches to temperature
measurement here, considering individual degrees of freedom within a nominally
one-dimensional shockwave in two-dimensional plane geometry.

Although the notion of configurational temperature can be defended at
equilibrium, the shockwave problem indicates a serious deficiency in
the concept.  An isolated row or column of particles, all with the same $y$
or $x$ coordinate and interacting with repulsive forces is clearly unstable to
transverse perturbations.  The symptom of
this instability is a negative ``force constant''
$$
(\langle \nabla \nabla {\cal H} \rangle )_{xx \ {\rm or} \ yy} < 0 \ .
$$
The uncertain sign of $\nabla \nabla  {\cal H}$ explains the presence of
wild fluctuations, and even negative configurational temperatures, in the
vicinity of the shockfront.  See Fig. 6. This outlandish behavior means
that the
configurational temperature is not a useful concept for such problems.

\begin{figure}
\includegraphics[height=16cm,width=8cm,angle=-90]{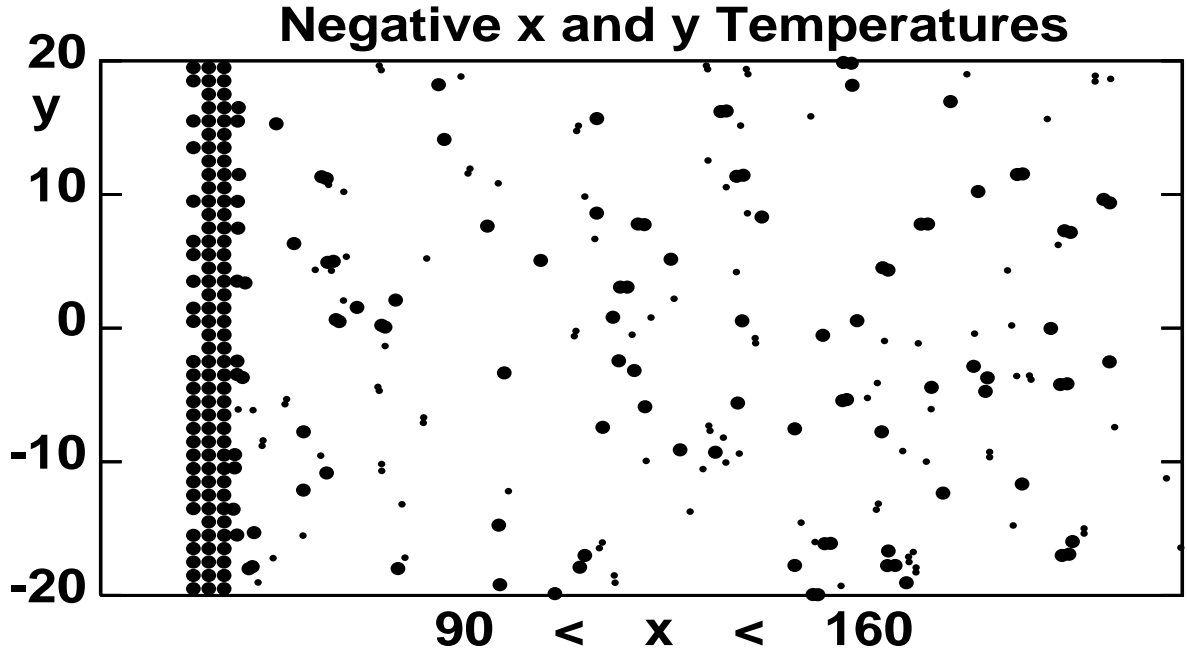}
\caption{
Particles with {\em negative} configurational temperatures are shown here,
using the data underlying Figs. 4 and 5.  The material to the right of the
shock is ``hot'', with cold material entering at the left.  The smaller
dots correspond to negative values of $T^{\Phi}_{xx}$ and the larger ones
to negative $T^{\Phi}_{yy}$.
}
\end{figure}

Defining the kinetic temperature requires first of all an average velocity,
about which the thermal fluctuations can be computed.  The velocity average
can be a one-dimensional sum over particles $\{ i\}$ sufficiently close
to the gridpoint $k$:
$$
u(x_k) \equiv \sum_i w_{ik}^{1D}v_i/ \sum_i w_{ik}^{1D} \ \longrightarrow T^{1D}
\ .
$$
Alternatively a local velocity can be defined at the location of each
particle $i$ by using a two-dimensional Lucy's weight function,
$$
w^{2D}(r<h) = (5/\pi h^2)[1 - 6(r/h)^2 + 8(r/h)^3 - 3 (r/h)^4] \ ; \ r < h=3 \ .
$$
$$
\longrightarrow \int_0^{h} 2\pi rw(r)dr \equiv 1 \ ,
$$
and summing over
nearby particles $\{ j\} $:
$$
u(x_i) \equiv \sum_j w_{ij}^{2D} v_j/ \sum_j w_{ij}^{2D} \ \longrightarrow T^{2D} \ .
$$
These latter two-body sums can either both include or both omit the
``self'' term with $i = j$.  Our numerical results support the intuition
that it is best to omit both the self terms\cite{b17}.  We compute all three
of these $x$ and $y$ temperature sets for the stationary shockwave profile and
plot the results in Figure 7 and 8.

\begin{figure}
\includegraphics[height=10cm,width=8cm,angle=-90]{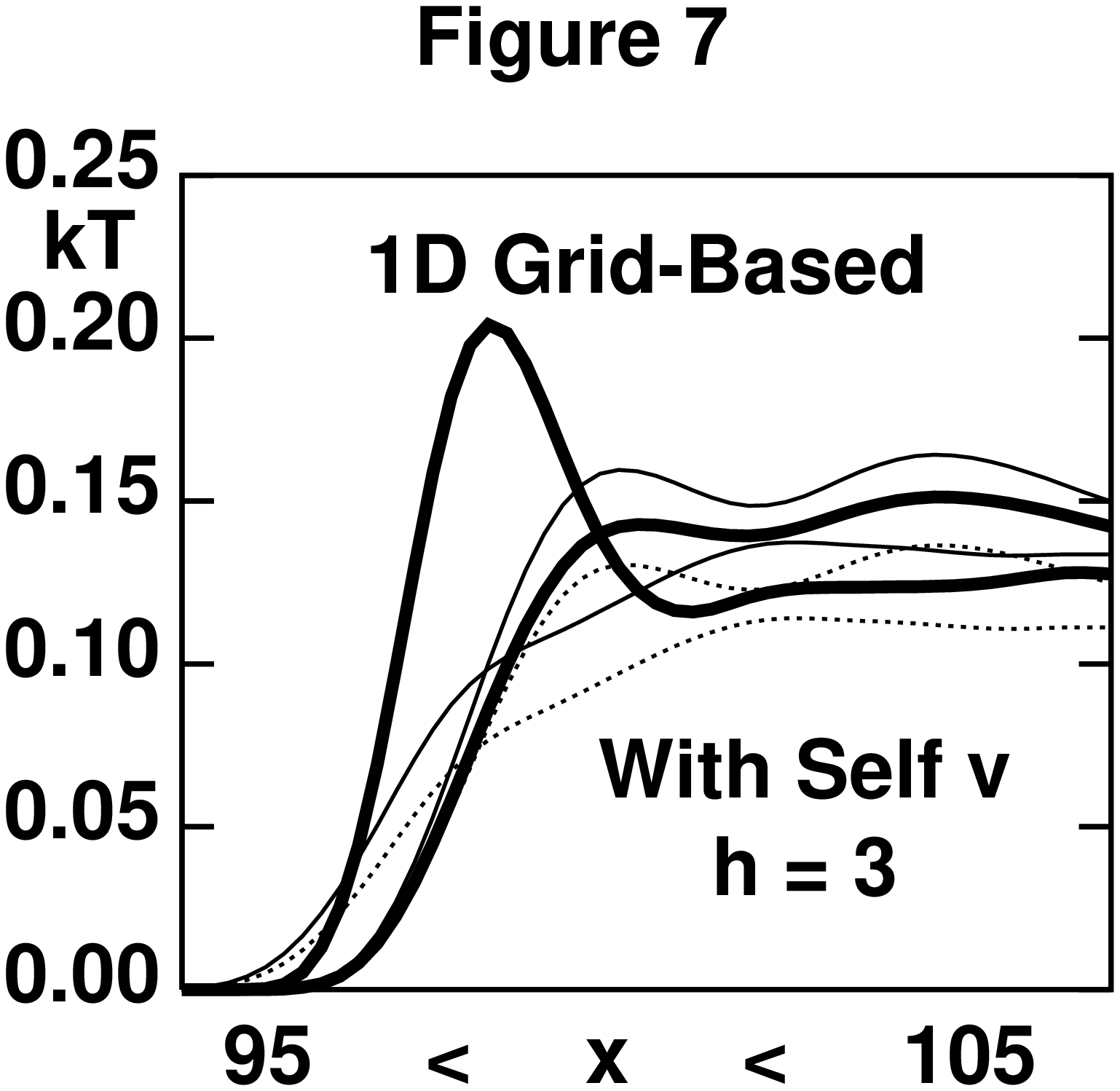}
\caption{
Typical instantaneous temperature profiles, at $t=5Dt$ for the strong
shockwave described in the text.  Local particle-based definitions of
stream velocity give the two somewhat lower temperature pairs,
$\ (T^K_{xx},T^K_{yy} \}$.  The grid-based definitions of $T^K_{xx}$ and
 $T^K_{yy}$ are indicated with the heaviest lines.  They both use a
one-dimensional weight function.  This definition gives a strong
temperature maximum for $T^K_{xx}$.  In all three cases the longitudinal
temperature $T^K_{xx}$ exceeds the transverse temperature $T^K_{yy}$ near
the shock front.  The temperatures obtained with two-dimensional weights
and including the ``self'' terms in the average velocity are indicated
by the dashed lines and are significantly lower than the rest 
throughout the hot fluid exiting the shockwave.  The ``correct'' hot
temperature, far from the shock, is $kT = 0.13$, based on separate equilibrium
molecular dynamics simulations.  The profiles shown here were all
computed from the instantaneous state of the system after a simulation
time of $5Dt = 200/u_s$.
}
\end{figure}

\begin{figure}
\includegraphics[height=10cm,width=8cm,angle=-90]{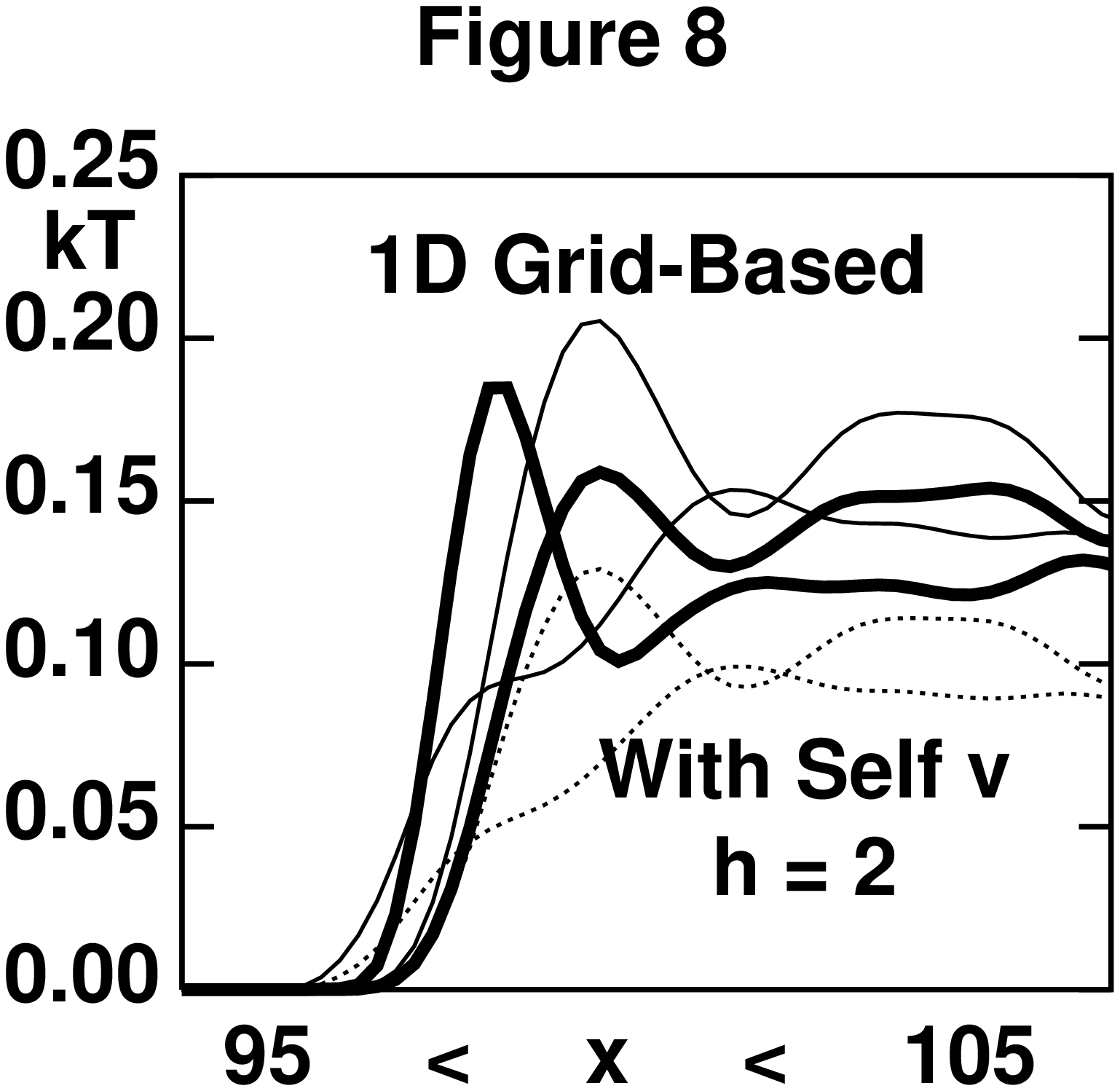}
\caption{
Temperature profiles with the same data as in Figure 7, but with a reduced
averaging range, $h=2$.  Again, the temperatures including ``self''
contributions to the local velocity are shown as dashed lines.  Notice the
sensitivity of the maximum in $T^K_{xx}$ to the range $h$. Far to the right,
the ``correct'' hot temperature, at equilibrium, away from the shock, is
$kT = 0.13$, based on separate equilibrium molecular dynamics simulations.
}
\end{figure}

The data show that in every case $T_{xx}$ exceeds $T_{yy}$ in the leading
edge of the shock
front. There is a brief time lag between the leading rise of the longitudinal
temperature $T_{xx}$ and the consequent rise of the transverse temperature
$T_{yy}$.  This is to be
expected from the nature of the shock process, which converts momentum in the
$x$ direction into heat\cite{b26,b27}. The Rayleigh line itself (see again
Figure 2) shows the mechanical analog of this anisotropy, with $P_{xx}$
greatly exceeding $P_{yy}$.  Note that the lower pressure in Figure 2,
the Hugoniot pressure, is a set of equilibrium values, $(P_{xx}+P_{yy})/2$.
It is noteworthy that the height of the temperature maximum is sensitive to
the definition of the local velocity.  Evidently a temperature based on the
local velocity, near the particle in question, and with a coarser averaging
range $h=3$, gives a smaller gradient
and should accordingly provide a much simpler modeling challenge.  

\section{Summary}

This work shows that planar shockwaves are stable for a smooth repulsive
potential in a dense fluid.  We also find that a mechanical instability
makes the configurational temperature quite useless for such problems.
Our investigation of temperature definitions shows that relatively smooth
instantaneous temperature profiles can be based on the weight functions
used in smooth particle applied mechanics.  It is noteworthy that the
constitutive relations describing inhomogeneous nonequilibrium systems
must necessarily include an averaging recipe for the constitutive properties.
The present work supports the idea\cite{b11} that the range of smooth
averages should be at least a few particle diameters.

The better behavior of a particle-based temperature when the ``self''
terms are left out is not magic.  Consider the equilibrium case of a
motionless fluid at kinetic temperature $T$.  The temperature of a
particle in such a system should be measured in a motionless frame.  If
the ``self'' velocity is included in determining the frame velocity an
unnecessary error will occur.  Thus it is plausible that the ``self''
terms should be left out.  The data shown in Figs. 7 and 8 support this
view.

The instantaneous particle-based velocity average provides a
smoother gentler profile 
which should be simpler to model.  By using an elliptical weight function,
much wider in the $y$ direction than the $x$, one could consider the grid-based
temperature as a limiting case.  The somewhat smoother behavior of the
particle-based
temperature recommends against taking this limit.  The elliptical weight
function would leave the divergent configurational temperature unchanged.

\section{Acknowledgment}

We thank Brad Lee Holian for insightful comments on an early version of
the manuscript.  The comments made by the two referees were helpful
in clarifying the temperature concept and its relation to earlier
work.  John Hardy was particularly forthcoming and generous in discussing
the early
history of his approach to making ``continuum predictions from molecular
dynamics simulations''.  This work was partially supported by the British
Engineering and Physical Sciences Research Council and was presented at
Warwick in the spring of 2009.


\begin{thebibliography}{99}

\bibitem{b1} G. E. Duvall and R. A. Graham, ``Phase Transitions under
             Shockwave Loading'', Reviews of Modern Physics {\bf 49},
             523-577 (1971).

\bibitem{b2} L. D. Landau and E. M. Lifshitz, {\em Fluid Mechanics}
             (Reed, Oxford, 2000).

\bibitem{b3} G. I. Kanel, W. F. Razorenov, and V. E. Fortov,
             {\em Shockwave Phenomena and the Properties of Condensed
             Matter} (Springer, Berlin, 2004).

\bibitem{b4} W. J. Nellis, ``P. W. Bridgman Contributions to the
             Foundations of Shock Compression of Condensed Matter''
             arXiv:0906.0106.

\bibitem{b5} R. E. Duff, W. H. Gust, E. B. Royce, M. Ross, A. C. Mitchell,
              R. N. Keeler, and W. G. Hoover, in {\em Behavior of Dense
              Media under High Dynamics Pressures}, Proceedings of the
              1967 Paris Conference (Gordon and Breach, New York, 1968).

\bibitem{b6} V. Y. Klimenko and A. N Dremin, in {\em Detonatsiya,
              Chernogolovka}, edited by G. N. Breusov {\it et alii}
             (Akademia Nauk, Moscow, 1978), page 79.

\bibitem{b7}  W. G. Hoover, ``Structure of a Shockwave Front in a Liquid'',
              Physical Review Letters {\bf 42}, 1531-1534 (1979).

\bibitem{b8} B. L. Holian, W. G. Hoover, B. Moran, and G. K. Straub, 
              ``Shockwave Structure {\em via} Nonequilibrium Molecular
             Dynamics and Navier-Stokes Continuum Mechanics'', Physical
             Review A {\bf 22}, 2798-2808 (1980).

\bibitem{b9} J. Hardy, ``Formulas for Determining Local Properties in
             Molecular-Dynamics Simulations: Shockwaves'', Journal of
             Chemical Physics {\bf 76}, 622-628  (1982), said by Hardy
             (private communication to WGH, June 2009) to be inspired
             by a 1963 conversation with Philippe Choquard (Lausanne)
             at the Lattice Dynamics Conference in Copenhagen.

\bibitem{b10} O. Kum, Wm. G. Hoover, and C. G. Hoover, ``Temperature
              Maxima in Stable Two-Dimensional Shock Waves'', Physical
              Review E {\bf 56}, 462-465 (1997).

\bibitem{b11} S. Root, R. J. Hardy, and D. R. Swanson, ``Continuum
             Predictions from Molecular Dynamics Simulations:
             Shockwaves'', Journal of Chemical Physics {\bf 118},
             3161-3165 (2003).

\bibitem{b12} C. E. Ragan III, ``Ultrahigh-Pressure Shockwave Experiments'',
              Physical Review A {\bf 21}, 458-463 (1980).

\bibitem{b13} H. M. Mott-Smith, ``The Solution of the Boltzmann Equation
             for a Shockwave'', Physical Review {\bf 82}, 885-892 (1951)

\bibitem{b14} O. G. Jepps, Ph. D. thesis, Australian National University
             (Canberra, 2001).

\bibitem{b15} C. Braga and K. P. Travis, ``A Configurational Temperature
              Nos\'e-Hoover Thermostat'', Journal of Chemical Physics
             {\bf 123} 134101 (2005).

\bibitem{b16}  Wm. G. Hoover and C. G. Hoover, ``Nonequilibrium Temperature
              and Thermometry in Heat-Conducting Phi-4 Models'', Physical
              Review E {\bf 77}, 041104 (2008).
    
\bibitem{b17} Wm. G. Hoover and C. G. Hoover, "Nonlinear Stresses and
              Temperatures in Transient Adiabatic and Shear Flows
              {\it via} Nonequilibrium Molecular Dynamics'', Physical
              Review E {\bf 79}, 046705 (2009).

\bibitem{b18}  B. D. Butler, G. Ayton, O. G. Jepps and D. J. Evans,
              ``Configurational Temperature: Verification of Monte Carlo
              Simulations'', Journal of Chemical Physics {\bf 109},
              6519-6522 (1998).

\bibitem{b19} Wm. G. Hoover, C. G. Hoover, and J. F. Lutsko, ``Microscopic
              and Macroscopic Stress with Gravitational and Rotational
              Forces, Physical Review E {\bf 79}, 0367098 (2009).

\bibitem{b20} L. B. Lucy, ``A Numerical Approach to the Testing of the
              Fission Hypothesis'',  The Astronomical Journal {\bf 82},
              1013-1024 (1977).

\bibitem{b21}  Wm. G. Hoover, {\em Smooth Particle Applied Mechanics ---
              The State of the Art} (World Scientific Publishers, Singapore,
              2006, available from the publisher at the publisher's site
              http://www.worldscibooks.com/mathematics/6218.html).

\bibitem{b22}  Wm. G. Hoover, {\em Computational Statistical Mechanics}
              (Elsevier, Amsterdam, 1991, available at the homepage
              http://williamhoover.info/book.pdf).

\bibitem{b23} W. G. Hoover, B. L. Holian, and H. A. Posch, ``Comment I on 
              `Possible Experiment to Check the Reality of a
              Nonequilibrium Temperature’$ \ $'', Physical Review E
              {\bf 48}, 3196-31998 (1993).

\bibitem{b24} Wm. G. Hoover and C. G. Hoover, ``Single-Speed Molecular
              Dynamics of Hard Parallel Squares and Cubes'', arXiv:0905.0293.
              (submitted, Journal of Statistical Physics, 2009).

\bibitem{b25} L. D. Landau and E. M. Lifshitz, {\em Statistical Physics}
              (Muir, Moscow, 1951)(in Russian), Eq. 33.14.

\bibitem{b26} B. L. Holian, C. W. Patterson, M. Mareschal, and E. Salomons,
              ``Modeling Shockwaves in an Ideal Gas: Going Beyond the
              Navier-Stokes Level'', Physical Review E {\bf 47},
              R24-R27 (1993).

\bibitem{b27} B. L. Holian, ``Modeling Shockwave Deformation {\em via}
              Molecular Dynamics'', Physical Review A {\bf 37}, 2562-2568
              (1988).


.



\end{thebibliography}
\end{document}